\documentclass[a4paper]{jpconf}
\usepackage{graphicx}
\begin{document}
\title{A functional model of interactions in quantum theory}

\author{ H H Diel}

\address{Diel Software Entwicklung und Beratung, Seestr.102, 71067 Sindelfingen, Germany}

\ead{diel@netic.de}

\begin{abstract}
A functional model of interactions in quantum
theory (QT) is proposed that describes the dynamic evolution
of a physical system in terms of process steps and intermediate
states; that is, it describes how things function.
The model of QT interactions has been developed in the context of the
author's work towards a computer model of QT with the goal of supporting the
widest possible scope of QT concepts. With standard QT, the results of interactions
are computed using quantum field theory (QFT). Therefore, the proposed
functional model of interactions must be compatible with QFT.
Described in terms of a cellular automaton, the model assumes discrete time, space and wave function paths.
\end{abstract}

\section{Introduction}

In 
\cite{Feynman1} 
R. Feynman writes: "I have pointed out these things because the more you see how strangely Nature behaves, the harder it is to make a model that explains how even the simplest phenomena actually work. So theoretical physics has given up on that."

"A model that explains how phenomena actually work" is referred to as  a "functional model" in this paper. 
A functional model of a given function explicitly describes a sequence of steps for the development of the function's final state. A more detailed explanation of what in this paper is understood as a "functional model"  is provided in \cite{diel2}.

The need for a functional model of QT was established when the author tried to develop a comprehensive computer model of QT.
He discovered that certain computability problems associated with QT/QFT (see \cite{diel1}) can be overcome with a functional model of QT/QFT. A second motivation for a functional model, in particular of QT interactions, was the author’s search for a solution to the QT measurement problem, one that did not require modifications of the Schr\"odinger equation (or similar equations of motion of QFT) or one that did not rely on assumptions outside the scope of falsifiable physics. In  \cite{diel4}, a model of a QT measurement is described where measurements are explained in terms of QFT interactions and their functional descriptions.

Quantum field theory (QFT) is the area of QT that addresses interactions among particles. The founders of QFT, including R. Feynman, recognized the need for a process-based theory of QT interactions. Feynman himself used the word "process" extensively when describing his quantum electrodynamics (QED) .
\footnote{ R. Feynman published a QED related book with the title "The Theory of Fundamental Processes" \cite{Feynman2}.} The most fundamental tool in QFT, Feynman diagrams, give the impression of showing a temporal structure, i.e., a process.
However, upon closer inspection, it turns out that the process orientation of QFT is insufficient and that QFT/QED must still be considered a declarative (i.e., non-functional)  description. The term "process" may be applicable when viewing a QFT interaction as a whole, but the theory does not provide any temporal substructures including intermediate states. 
Feynman diagrams must not be viewed as showing a temporal structure. The fact that multiple diagrams may have to be applied for a (single) scattering process disturbs any temporal interpretation. Additionally, possible loops in Feynman diagrams do not mean temporal loops. 


The description of (physical) processes requires a certain specification language. Only very simple dynamic processes can be specified purely in terms of mathematical equations such as differential equations. Unfortunately, there does not exist a generally agreed process (i.e., functional) specification language that is comparably as powerful and compact as the language of mathematical equations used in physics for the description of static relationships (including derivatives). The method of description used in this paper is a mixture of plain English, some semi-formal algorithmic specification language, and the mathematical equations of QFT.

The model is based on cellular automaton (CA). A CA implies a special structure for the overall dynamic evolution and for space-time. The most important implication is the discreteness with respect to time and space.

This functional model of interactions in QT is embedded in the overall functional description of QT described in 
\cite{diel2}.
The key features of this overall functional description of QT are listed in Section 2. A functional description of a system describes the dynamical evolution of this system in terms of state changes. Therefore, in Section 3, the structure and components of the system state of the functional model is first described before the steps in interaction processes are described in Sections 5 and 6.
There are a number of important QT concepts, such as measurement, entanglement, and decoherence, that are related to interactions. These subjects are also addressed in the present paper in section 7.

In this paper, interactions are assumed to always occur between two "in" particles and to result in two "out"  particles (not necessarily identical to the "in" particles). Interactions between more than two particles are assumed to be  translatable into a series of interactions between two particles. Special cases, for example,  those in which QFT allows for more than two "out" particles, will not be addressed. 



\section {Key Features of the Overall Functional Description of QT}

The functional description of interactions in QT, described in this paper, is embedded in the overall QT functional description  
\cite{diel2}.
The following  features are of primary importance for the overall functional description of QT. 

\subsection{Discreteness of QT attributes}

In 
\cite{tHooft} 
G. 't Hooft writes "Often, authors forget to mention the first, very important, step in this logical procedure: replace the classical procedure one wishes to quantize by a strictly finite theory. Assuming that physical structures smaller than a certain size will not be important for our considerations, we replace the continuum of three-dimensional space by a discrete but dense lattice of points. ". 

The functional description of QT assumes discrete and coarse grain attributes 
not only for  three-dimensional space, but for most other entities where standard QFT assumes differentiable attributes.
This applies to the spatial extension of particles/waves  and to  their momentum. Also, the wave function is structured into a
discrete set of alternative paths.

Of course, the graining has to be kept fine enough to prevent significant deviations from the predictions obtained
with standard QFT.

\subsection{The Transition from Possibilities to Facts - Handling of Non-Determinism}

The functional description/model of QT interactions must demonstrate the evolution of the wave function to generate probability amplitudes in accordance with the predictions of QT/QFT. However, it does not end with the determination of probability amplitudes but includes
a model for the realization of the predictions represented by the probability amplitude. This process step  is called
"the transition from possibilities to facts". With standard QT, the transition from possibilities to facts is a non-deterministic process step that occurs exclusively with measurements. 

One of the key features of the functional description of QT  is that  the transition from possibilities to facts  is not exclusively tied to measurements. With most interpretations of QT,  the measurement process implies a "collapse of the wave function".
There is an ongoing debate among QT physicists whether a collapse of the wave function has to be assumed. 
The functional description assumes that measurements always imply interactions, more specifically, interactions that lead to a collapse of the wave function. The statement that  the transition from possibilities to facts  is not exclusively tied to measurements, first of all, means that the collapse of the wave function is not exclusively observed with measurement interactions. The collapse of the wave function also occurs with other ("normal") interactions.

\subsection{Particle Fluctuations and Interaction Channels Instead of Virtual Particles}

In the perturbation (Feynman) approach, virtual particles are an essential concept for describing interactions among
particles. 
The functional description reinterprets the role of virtual particles; instead of the original QFT virtual particles, the functional description assumes “particle fluctuations" and "interaction channels". Particle fluctuations initiate the interaction, if the fluctuation affects multiple (i.e., at least two) particles. These particle fluctuations are assumed to actually occur (with a certain probability), whereas virtual particles are constructs that affect only the probability amplitudes.
Interaction channels (like those mediated by virtual particles) guide the possible flow of particle transitions during an interaction (see Section 6.2).

\subsection{Splitting of a Wave Function \emph{Collection} into Multiple Paths}

The splitting of a wave function into multiple paths is a constituent part of the perturbation approach to QFT (see  \cite{Feynman1}). 
The overall effect of the wave function progression is then determined by the superposition (via path integrals) of the multiple paths.
With the QT functional description, the splitting into multiple paths is applied to \textbf{collections} of particles 
which exit an interaction.
This allows for the modeling of entanglement (see section 7.6).

\section{The System State} 

Because the QT functional model does not distinguish between a particle and the (associated) wave, the  term "particle/wave" will be used in the following.
\footnote{In the literature on QT some authors used the name "wavicle" for what here is called "particle/wave".}

The description of the evolution of a quantum system, such as a collection of
particles/waves, must be related to the information that makes up this quantum
system. 
For the functional model it is useful to arrange the information in a certain structure, primarily derived from  the  differing variability and lifetime of the entities. 
The information that represents the quantum system for the functional
model must encompass all of the entities known from QT/QFT, such as state
vectors, wave functions, masses, charges, etc. 
In addition, the functional model must include objects and state components which are suited for the description of intermediate states. 

For the functional model, the totality of information constituting a quantum
system consists of a set of q-objects plus a set of fields.
\small{
\begin{verbatim}
QT-system :=   
  q-object-set,
  field-set;
\end{verbatim}
}
\normalsize
Fields (i.e., "field-set") are not further addressed in the present paper.

\subsection{The Quantum Object, q-object} 

The most general entity for the description of a quantum system is the q-object.
A q-object is an aggregate object that can be described by a common wave function and not just the product of the wave functions of the elements of the q-object. A particle/wave may occur as a separate q-object, or may be part of a q-object.
\\
For example, the wave function 

(1) $  \psi =   1/ \sqrt{2}\;  ( | \; pw1.up, pw2.down > + \;  | \;  pw1.down, pw2.up) >   $ 
\\
refers to a q-object $  \psi $ with elements (e.g., entangled particle/waves) pw1 and pw2.

A q-object may be viewed as having a two-dimensional structure. One dimension represents the elements of the q-object (with the above example, pw1 and pw2); the other dimension represents alternatives that may be selected during the evolution of the q-object, for example, by an interaction. In this paper, these alternatives are called "paths". Each path has associated a probability amplitude.

\small{
\begin{verbatim}
q-object :=   
  path[1],
  ...
  path[NPATH];
\end{verbatim}
}
\normalsize
\begin{table}
\caption{\label{label}Structure of a q-object consisting of two particle/waves pw1 and pw2.}
\begin{tabular} { | c | c | c  | c | }
\hline
paths & pw1-state & pw2-state  & amplitude  \\

\hline

path-1	  &   pw1-state$_{1}  $    &  pw2-state$_{1} $  & ampl-1  \\

path-2	 &   pw1-state$_{2}  $    &  pw2-state$_{2} $   & ampl-2  \\

...	            & ...         & ...    & ...  \\

path-N	 &  pw1-state$_{N} $    &  pw2-state$_{N} $  & ampl-N  \\
\hline
\end{tabular}
\end{table}  
\small{
\begin{verbatim}
path :=   
  state-element[1], ...,state-element[n], amplitude;
\end{verbatim}
}
\normalsize
With the above example at least two paths, path[1] := ( pw1.up, pw2.down ) and 
path[2] := ( pw1.down, pw2.up), both with amplitude =  $ 1/ \sqrt{2} $ may express the state of a q-object with a specific system evolution. 

The wave function for $ \psi $, eq. (1), specifies a continuous set of possible measurement results for  $ \psi $. Conversely, the q-object of the functional model  specifies a discrete set of paths that may be selected with interactions and measurements.
 
Different types of q-objects, containing different types of elements, can be distinguished.
The simplest type of q-object is the (single) particle/wave. In this paper, in addition to the particle/wave, two other q-objects, the particle/wave-collection (pw-collection) and the interaction-object, are of particular importance. A pw-collection represents a collection of particles/waves that are entangled. Table 1 shows the structure of a pw-collection.
The interaction object is an interaction internal object created at the beginning of an interaction. At the end of the interaction, the interaction object is transformed to a pw-collection.

In addition to the q-objects addressed in this paper, there are further entities in QT, such as bound systems, which may fall under the concept of q-objects. This functional model assumes that q-objects are created from interactions and dissipate (i.e., collapse) when they get involved in new interactions.
\footnote{As a consequence, the idea of the whole universe constituting a single big q-object (with a common probability amplitude) is not supported by the functional model.}

The state of an element of a q-object (above denoted state-element []) consists of the state-components known from QFT.
For QED, these are, first of all, the parameters used in QFT to specify a matrix element of the scattering matrix
$ \Psi_{p1,\sigma1,n1;p2,\sigma2,n2, ...} $, i.e., the four-momenta $ p^{\mu}$, the spin z-component (or for massless particles, helicity) $ \sigma $, and the particle type n. The Lagrangian, Hamiltonian, and the equations of motion for the typical fields of QFT contain the time derivative $ d\phi / \partial t $ of the state $ \phi $. For the functional model, this time derivative therefore  has to be included as part of the state. 
In addition, the position vector x is part of the state.

\section{The Cellular Automaton}

To describe the dynamical evolution of the states of a system, some language or description method is required. 
The method chosen in this paper for the specification of the dynamical evolution of 
a QT systems a cellular automaton (CA).

\subsection{Standard Cellular Automaton}
The standard CA consists of a k-dimensional grid of cells. The state of the CA is given by the totality  of the states of the individual cells.

   $    S_{CA} =  \{ s_{1}, ... , s_{n} \} $

With traditional standard CAs, the cell states uniformly consist of the same state components

   $    s_{i} =  \{ s^{1}_{i}, ... , s^{j}_{i} \} $
\\
Typically, the number of state components, j, is 1, and the possible values are restricted to integer numbers.

The dynamical evolution of the CA is given by the (single general) "update-function" which computes the new 
state of a cell as a function of  its current state and the states of the neighbor cells.
\small{
\begin{verbatim}
Standard-CellularAutomaton(initial-state)  :=     // transition function
DO FOREVER {
   state = update-function(state, timestep);
   IF ( termination-state)  STOP;
}
\end{verbatim}
}
\normalsize
The full complexity (if any) of a particular cellular automaton  is concentrated in the update-function. As Wolfram (see \cite{Wolfram}) and others (see, for example, \cite{Ilachinski}) showed, a large variety of process types (stable, chaotic, pseudo-random, oscillating) can be achieved with relatively simple update-functions.

For specific applications of the cellular automaton, the update-function may be derived from application specific specifications.
In \cite{Elze1}, Elze describes a cellular automaton whose update-function is derived from the Hamiltonian (or the equation of motion). 

\subsection{QFTCA, a Cellular Automaton Supporting QFT }

The CA described in \cite{Elze1} , \cite{Elze2}, and  \cite{Hooft} may be viewed as the starting points for the CA described in the present paper. 
To support QFT (at least to the extent required to demonstrate a model of interactions), a cellular automaton QFTCA is defined; QFTCA, however, requires certain extensions of the standard cellular automaton and embedding of the cellular automaton in an overall QFT-based structure. The embedding into the QFT-based structure affects two aspects, the structure of the state of the system and the functions for the dynamic evolution of the system.

\subsubsection{State structure}

The CA-cells may be viewed as representing the space.
The cells have associated with them (i.e., contain) parts of the system state.
For the mapping of the overall system state, as described in Section 3, to QFTCA, some "design decisions" have to made.
The following lists the major characteristics (without supporting arguments):
\begin{itemize}
\item Time is \emph{not} a (fourth) dimension for QFTCA. Instead, the time derivatives  $  d\psi / \partial t $ and $ p^{\mu}$ are explicit parts of the system state.
\item Differing from the standard CA (mentioned in Section 4.1), the states of QFTCA cannot be expressed solely by the collection of cell states. Those parts of the q-object state that are position dependent can be assigned to cell states. Other parts have to be kept besides the cell states. For the functional model described in this paper, left open is the assignment of state components to CA cells as opposed to state components that are kept with the (paths of the) q-objects. We need to ensure that the cells (i.e., space points) belonging to a q-object can be determined and vice versa, that the q-objects occupying a CA cell can be determined.
\item A q-object path may cover multiple cells.
\item A cell may be covered by multiple q-object paths.
\item The above-described state structure and timing considerations result in differing content of CA cells and differing CA update functions. For the proposed model of the QT measurement process, this could be implemented by either  (a) a single relatively complex QFTCA, or (b) by a QFTCA consisting of a collection of  CAs that will partly merge whenever interactions are performed.
In the present paper, the decision of (a) or (b) and the details of (a) and (b) are left open.
\end{itemize}

\subsubsection{Evolution of the system state} 
The structuring of the overall system state into  the CA-cells and  the superimposed q-objects was also motivated by the differing update requirements of the QT-objects.
\small{
\begin{verbatim}
QFTCA (initial-state)  :=  // transition function
DO FOREVER {
   state = global-update-function(state, timestep);
   IF ( termination-state)  STOP;
}

global-update-function(state, timestep)  :=    
DO PARALLEL {
  field-state = field-update-function(field-state, timestep);
  FOR ( all qobjects qobj[i] ) {
        propertimestep = fx(qobj[i]) * timestep;
        FOR ( all particle/waves pw[k] of qobj[i]) {
              pw[k] = pw-update-function(pw[k], propertimestep);
             IF ( interaction-occurred( pw[k], pw2 )  
                   perform-interaction( pw[k], pw2 );	
        }
  }
}
\end{verbatim}
}
\normalsize
QFT is a relativistic theory. Special relativity distinguishes proper time (or wrist-watch time) of the inertial system and global time of the overall space-time system.
For the evolution of the overall system (state), the functional model assumes that the QFTCA proceeds in uniform global time steps. The update-function for the individual q-objects, however, have to proceed in proper time associated with the 
q-objects.

As a simplified description "propertimestep = fx(qobj[i]) * timestep;" indicates the transition from global time to proper time as requested by special relativity. A more detailed and more proper description is outside the scope of this paper.

The update functions for fields (field-update-function()) and for the normal particle propagation (pw-update-function()) are straightforward and not topics discussed in this paper. The topics discussed are the update functions related to interactions (interaction-occurred() and perform-interaction()). These are addressed in Sections 5 and 6.

\section{Overall Model of Interactions between Particles/Waves}

Two types of interaction between particles/waves are distinguished: (1) interactions
that destroy the superposition among possible multiple paths of a wave function, resulting in a collapse of the wave function, and (2) interactions that only affect the attributes (e.g., momentum) of the involved particles/waves. For this description of the functional model, interactions resulting in a collapse of the wave function are considered to be the general case. Interactions that do not result in a collapse are exceptions. They are called volatile interactions and will be addressed where appropriate and explicitly in Section 7.2.

With QFT (see \cite{Weinberg}) an interaction (e.g., scattering) is described by the scattering  matrix (S-matrix)  which assign a probability amplitude $ S_{\beta \alpha}$ to  the transition of a given "in"  state $ \Psi_{\alpha} $ to an "out" state $ \Psi_{\beta} $.
\\
 $ S_{\beta \alpha} = ( \Psi_{\beta}, \Psi_{\alpha} )  $
\\
The "in" and "out" states are specified by their state components (see Section 3)
  $  \Psi_{\alpha} = \Psi_{p1,\sigma1,n1; p2,\sigma2,n2, ...};  \Psi_{\beta} = \Psi_{p1',\sigma1',n1'; p2',\sigma2',n2', ...}  $
\\
The functional model of QFT interactions has to provide a model for the process that transforms an "in" state
$  \Psi_{\alpha} = \Psi_{p1,\sigma1,n1; p2,\sigma2,n2, ...}$ into a multitude of possible "out" states $ { \Psi_{\beta 1}, 
\Psi_{\beta 2}, ... }$ .
This multitude of possible "out" states constitutes a pw-collection as presented in Table 1.

The overall model of interactions between particles/waves is based on the following assumptions:
\begin{itemize}
\item Interactions are process steps that actually occur (with a certain probability) rather than wave function alternatives which are in superposition with "no interaction occurrence".
\item An interaction always occurs at a definite (discrete) point in space-time.
\\
(The QFT model of interactions in coordinate space, where the possible results of an interaction are computed by assuming superpositions among all possible interaction positions, is not supported by the functional model.) 
\item Only those q-object paths that cover  the interaction position affect the outcome of the interaction. If multiple paths of a q-object cover the interaction position, only one of them is 
selected as the interacting path.
\item The non-selected paths are discarded.
\item The selection of the interaction position and of the significant path represents (a first step in) the transition of probabilities to facts.
\item The discarding  of the non-selected paths can be viewed as "the collapse of the wave function".
\end{itemize}
\begin{figure}[ht]
\center{\includegraphics*[scale=0.5] {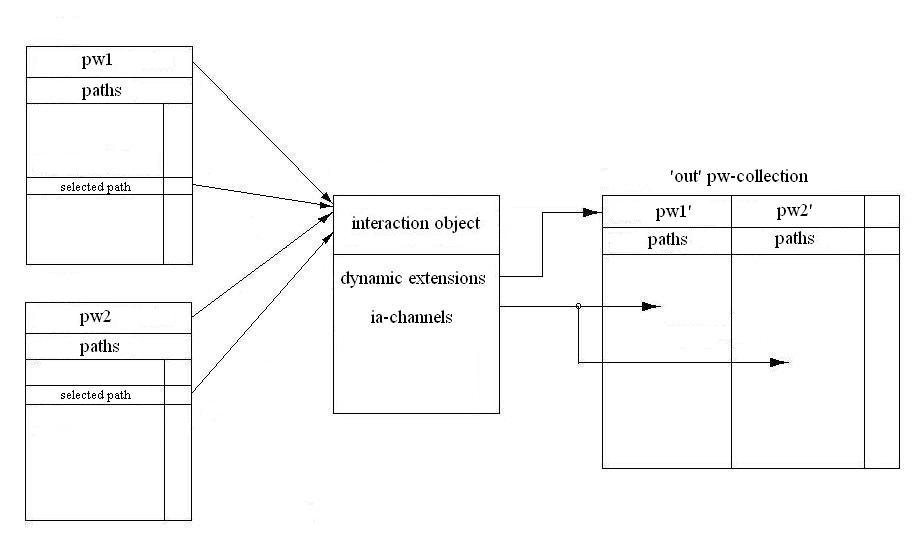} }
\caption{Information flow with the interaction between two particle/waves}
\end{figure}
Figure 1 shows the overall flow of information with the interaction of two particle/waves pw1 and pw2. The figure contains four q-objects: pw1, pw2 (the figure does not assume pw1 and pw2 belonging to the same q-object), the interaction-object, and the "out" pw-collection containing the "out" particle/waves.
At the beginning of the interaction the information contained in pw1 and pw2 is merged into the interaction object. The 
interaction object performs a sophisticated process, including the formation,  union, and splitting of "ia-channels", which finally results in the "out" pw-collection.

\subsection{Occurrence of an interaction}

Although QFT provides precise rules for the computation of probabilities (amplitudes) for the occurrence of specific interaction results as a function of the "in" states of the involved particle waves, it does not provide further details on the circumstances that must hold for an interaction to occur. In contrast, the functional model of QT interactions has to include a process model for the occurrence of an interaction that supports the above listed basic assumptions. The author claims that measurements imply interactions and therefore interactions occur with probabilities that are equal to the probabilities from the results of measurement.
The model chosen by the author for the occurrence of  interactions is the "particle/wave fluctuation" (pw-fluctuation).
A pw-fluctuation can be thought of as a temporary concentration and amplification of one or several particles/waves at a certain point in space. The following assumptions are essential in considering pw-fluctuations and their role in the functional model of QT interactions:
\begin{itemize}
\item With the functional model each interaction is preceded by a pw-fluctuation. 
\item Only one pw-fluctuation can be active at a given point in time for a  particle/wave.
\item The position where the pw-fluctuation occurs can be anywhere within the space occupied by the involved
particles/waves. The position is determined
randomly as a function of the amplitudes of the involved particles/waves and of the fields involved.
\end{itemize}
The immediate effect of a pw-fluctuation is the temporary formation of an entity called an interaction-object (see section 6.1). 
When the temporary interaction object disappears again, the original particles/waves may persist (or
may be reinstalled), or a different set of particles/waves may appear. Accordingly, the
long-term effect of a pw-fluctuation can be one of the following:
\begin{enumerate}
\item nothing durable (this may be the case with the majority of pw-fluctuations),
\item an interaction \emph{with} a collapse of the wave function (the major case considered in this paper),
\item an interaction \emph{without} a collapse of the wave function (only roughly addressed in section 7.2),
\item a particle decay (not further addressed in this paper).
\end{enumerate} 

\subsection{Results of an interaction}
This functional model of QT interactions is based on QFT. With QFT, the possible results of an interaction are provided by the S-matrix. It includes probability amplitudes for possible results for a given "in" state. The functional model has to deliver the collection of the complete variety of possible "out" states and their probability amplitudes in form of the "out" pw-collection.

\section{Functional Steps in an Interaction Process}

The functional description must show process steps for the dynamic evolution of an interaction. The process steps have to produce results that are compatible with the results predicted by standard QT/QFT in form of the S-matrix.

First thoughts on the mapping of QFT to a functional model may start with looking at the operators and operator equations of QFT. For example, in \cite{Mandl} the "\emph{processes}" that contribute in QED to the determination of the S-matrix, are expressed as

$ H_{W}(x) = -eN \{ ( \bar{\psi^{+} } +  \bar{\psi^{-} }) ( \not A^{+} + \not A^{-}) ( \psi^{+} -  \psi^{-}) \}_{x} $
\\
where $ \bar{ \psi^{+}}, \bar{ \psi^{-}}, \not A^{+}, \not A^{-},  \psi^{+},  \psi^{-} $ are creation and annihilation operators.

The further QFT treatment of interactions leads to Feynman diagrams. Although Feynman diagrams already seem to contain certain process-oriented aspects, the QFT operators and Feynman diagrams are not directly usable as  basis for a functional (i.e., process-based) description. 

Instead of the QFT creation and annihilation operators, the functional model contains split() and combine()  operators. Similar to the creation and annihilation operators, the split() and combine()  operators have to appear in pairs.
Instead of the Feynman diagrams, the functional model contains "interaction channels". 
Of course, it has to be ensured that the functional model results in QFT compatible predictions.

The description of the overall interaction process is subdivided into three process steps:
\small{
\begin{verbatim}
perform-interaction ::= {
  Step1: Formation of interaction-object;
  Step2: Formation and processing of ia-channels;
  Step3: Generation of "out" particle/wave collection;
}
\end{verbatim}
}
\normalsize

\subsubsection{Example: Bhabha Scattering}
QFT provides rules and equations for the computation of scattering matrix amplitudes. The equations are derived from the pertinent Feynman diagrams. For Bhabha scattering the equations are

(1)  $ M_{A} =  (-ie)^{2} \bar{v}(\vec{p}_{2}, s_{2} ) \gamma_{\mu}  u(\vec{p}_{1}, s_{1} ) (-ig^{\mu\nu}/(p_{1} + p_{2})^{2})
 \bar{u}(\vec{p}'_{1}, s'_{1} ) \gamma_{\nu}  v(\vec{p}'_{2}, s'_{2} ) $.
\\and

 (2)  $ M_{B} =  (-ie)^{2} \bar{u}(\vec{p}'_{1}, s'_{1} ) \gamma_{\mu}  u(\vec{p}_{1}, s_{1} ) ( -ig^{\mu\nu}/(p_{1}-p'_{1})^{2})
 \bar{v}(\vec{p}_{2}, s_{2} ) \gamma_{\nu}  v(\vec{p}'_{2}, s'_{2} ) $
 \\
According to the usual QFT notation $ u() $  represents the "in" electron, $ \bar{u}()$  the "out" electron, $ \bar{v}$  the "in" positron, and $ v() $ the "out" positron. $ M_{A} $ and $ M_{B} $  are the probability amplitudes. The total probability amplitude M for  Bhabha scattering (first order perturbation) is $ M = M_{A} - M_{B} $.

The "in" electron  $ u() $ and the positron $ \bar{v}() $ provide the initial state of pw1 and pw2. Differing from standard QFT computations, no specific value can be assumed for the states of  $ \bar{u}()$ and $ v() $ when an interaction starts within the functional model. Therefore, the result of an interaction within the functional model will not be a single probability amplitude M, but a set of probability amplitudes embraced in the "out" pw-collection. 

\subsection{Formation of interaction object}

At the beginning of an actual interaction, the information from the interacting particles/waves is combined into the interaction object. At the end of the interaction, the interaction object is replaced by a new particle/wave collection
representing the "out" particles/waves. The interaction object is a special type of q-object.
When the interaction object is initialized, it contains only a single path which contains the information from the selected paths of the two "in" particles/waves.
\small{
\begin{verbatim}
pw-ia-object :=   path[1];

path :=
  pw1.selectedPath.attributes,   pw2.selectedPath.attributes,  amplitude;
\end{verbatim}
}
\normalsize
To support the transition from the "in"
particles/waves to the "out" particle/wave collection, the interaction object has
a very dynamic structure.
During the interaction process, the interaction object will be extended by ia-channels which reflect 
intermediate states.
\small{
\begin{verbatim}
pw-ia-object :=   
  path[1] := ia-channel[1],
  ...
  path[k] := ia-channel[k];
\end{verbatim}
}
\normalsize

The functional model assumes that interaction objects, similarly to virtual
particles, have a limited life-time before they decay into the particles/wave collections that
are the result of the interaction. 

\subsubsection{Example: Bhabha Scattering}
For the formation of the interaction object the mapping of QFT to the functional model is still rather trivial. $u(\vec{p}_{1}, s_{1} ) $ is mapped to pw1.selectedPath.attributes;  $  \bar{v}(\vec{p}_{2}, s_{2} ) $ is mapped to   pw2.selectedPath.attributes. Differing from standard QFT, however, the functional model assumes only single paths to be selected as input for the further processing of the interaction. 

\subsection{Formation and processing of ia-channels}

The processing of interactions proceeds with the formation of ia-channels. Like the paths of a pw-collection (see section 3), an ia-channel represents one or multiple alternative(s) for the evolution of the "in" particles/waves toward the
interaction result. Each ia-channel starts with both "in" particles/waves and
ends with "out" particles/waves. Alternative ia-channels may differ in the set
of "out" particles/waves and/or in the sub-channels between the "in" and "out"
particles/waves.

A specific ia-channel is formed by the combination of the two operators
splitl() and combine().

 split(a) $ \rightarrow $  (b,c)  means   that split(a) results in b and c;

combine(a,b)$  \rightarrow $  (c)  means that combine(a,b) results in c.
\\
For example, starting with two interacting particles/waves pw1 and pw2, the ia-channel  
\\
  (pw1,pw2): split(pw1) $  \rightarrow $ (a,b);  combine( a, pw2 ) $ \rightarrow $ ( c)   
\\ would result in "out" particles/waves b and c.

The split() and combine() operators are analogous (although not equal)  to the creation and annihilation operators of QFT. 

Derived from QFT the following rules are established for the functional model of interactions:
\begin{itemize}
\item Rule1: An interaction always starts with two "in" particles/waves and ends with two "out" particles/waves,
\item Rule2: An ia-channel always contains one combine() and one split() (in arbitrary sequence).
\footnote{In QFT deviations from these rules can be found which, however, will not be addressed here.} 
\end{itemize}
Given the above rules, there are five possible ia-channels which can be constructed from the two interacting particles/waves pw1 and pw2:
\begin{enumerate}
\item combine( pw1, pw2 )   $  \rightarrow $ (a)  split(a)  $  \rightarrow $ (b,c)
\item split(pw1)  $  \rightarrow $ (a,b) combine( a, pw2 ) $  \rightarrow $ (c)
\item split(pw1)  $  \rightarrow $  (a,b) combine( b, pw2 ) $  \rightarrow $ (c)
\item split(pw2)  $  \rightarrow $  (a,b) combine( pw1, a ) $  \rightarrow $ (c)
\item split(pw2)  $  \rightarrow $  (a,b) combine( pw1, b ) $  \rightarrow $ (c)
\end{enumerate}
All these ia-channels end with two particles/waves (a,c) or (b,c). Depending on
the type of particles/waves involved in the interaction, QFT supports only specific
kinds of splits and combines. The rules that define what combinations of particle
types may be subject to combine(p1, p2) and what the resulting particle types
of split(p1) and combine(p1, p2 ) can be are equivalent to the rules regarding the
possible vertices of Feynman diagrams as described in numerous textbooks on
QFT
 (see for example, \cite{Ryder}, \cite{Griffiths},  \cite{Mandl}).
 For QED, for example, one of the rules is
 split(  $ \gamma) \rightarrow  ( e^{-},  e^{+}) $;   combine($ e^{-}, \gamma) \rightarrow ( e^{-}) $ .
\\
Sometimes, QFT rules allow for multiple results for split() and combine(). For example, split(photon) may result in (electron, positron), (muon, antimuon),
or (tauon, antitauon). Finally, with specific "in" particle/wave combinations, it may turn out that some of the possible
 ia-channels may be considered equivalent, and therefore, only one
of them has to be included. 
\footnote{The rules regarding when ia-channels may be considered equivalent are defined with QFT (in terms of equivalent Feynman diagrams) and are not further addressed here.}
Given the QFT rules, typically only one or two of the above mentioned five alternative ia-channels are possible for a specific "in" particle combination.

The operator split(a) $\rightarrow $ (b, c) is a non-bijective function insofar as there
are many alternatives with respect to the attributes (e.g., momentum and spin)
of the resulting (b, c). Rather than selecting a particular specific result, QFT
(and the functional model of QT interactions) requires that the multitude of
possible results is generated, with differing probability amplitudes assigned.  
\footnote{The functional model of QT interactions assumes a certain granularity with respect to the multitude of possible results.}
Thus,   split(a) $ \rightarrow $ (b, c) results in a two-fold splitting and may be expressed as

split(a) $ \rightarrow  ((b_{1}, c_{1}),  (b_{2}, c_{2}), ... (b_{n}, c_{n})) $. 
\\
For the interaction object this means that a multitude of paths representing $  ((b_{1}, c_{1}),  (b_{2}, c_{2}), ... (b_{n}, c_{n})) $ has to be generated. As a consequence, at the end,  each ia-channel contains a multitude of paths, because each ia-channel includes a split() operator.
The alternative ia-channels that are generated by the varying application of 
the split() and combine() operators are processed in parallel.  The effects
of processing the split() and combine() operators is reflected in extensions of and
changes in the interaction object. 

The rules governing the computation of the amplitudes of a path of the interaction object must be in accordance with QFT. These rules are well known and described in many textbooks
on QFT
(see for example, \cite{Ryder}, \cite{Griffiths},  \cite{Mandl}). 
However,
with QFT the respective rules are defined in terms of (external and internal ) lines and vertices of Feynman diagrams. For the functional model, the QFT rules have been mapped to rules regarding the split() and combine() operators. 
\footnote{This mapping is not described in this article.} 

\subsubsection{Example: Bhabha Scattering}
Bhabha scattering refers to the electron-positron scattering process  $ ( e^{-}, e^{+} ) \rightarrow ( e^{-}, e^{+} ) $.
Derived from the rules of QFT (more specifically, quantum electrodynamics), the following ia-channels are possible:
\begin{itemize}
\item CA: $ combine1( e^{-}, e^{+} ) \rightarrow ( \gamma );$      $   split2( \gamma ) \rightarrow (  e^{-}, e^{+} ) $
\item CB1:  $ split1( e^{-}) \rightarrow (  e^{-}, \gamma );  $   $ combine2( \gamma,  e^{+} ) \rightarrow ( e^{+} )$
\item CB2: $ split1( e^{+}) \rightarrow (  e^{+}, \gamma ); $    $ combine2( \gamma,  e^{-} ) \rightarrow ( e^{-} )$
\end{itemize}
CB1 and CB2 can be shown to be equivalent. Therefore in the following only CA and CB (=CB1) will be considered. As can be easily demonstrated CA corresponds to equation (1) for $ M_{A} $ whereas CB  corresponds to equation (2) for $ M_{B} $.
\\
Consequently, CA $ = combine( e^{-}, e^{+} ) \rightarrow ( \gamma );   split( \gamma ) \rightarrow (  e^{-}, e^{+} ) $ has to be mapped to \\
 $  (-ie)^{2} \bar{v}(\vec{p}_{2}, s_{2} ) \gamma_{\mu}  u(\vec{p}_{1}, s_{1} ) (-ig^{\mu\nu}/(p_{1} + p_{2})^{2})
 \bar{u}(\vec{p}'_{1}, s'_{1} ) \gamma_{\nu}  v(\vec{p}'_{2}, s'_{2} )  $ and
\\
CB $ = split( e^{-}) \rightarrow (  e^{-}, \gamma ), combine( \gamma,  e^{+} ) \rightarrow ( e^{+} )  $ has to be mapped to \\
 $ (-ie)^{2} \bar{u}(\vec{p}'_{1}, s'_{1} ) \gamma_{\mu}  u(\vec{p}_{1}, s_{1} ) ( -ig^{\mu\nu}/(p_{1}-p'_{1})^{2})
 \bar{v}(\vec{p}_{2}, s_{2} ) \gamma_{\nu}  v(\vec{p}'_{2}, s'_{2} ) $.
\\
This allows the derivation of the general function logic for the split() and combine() functions.Without going into further details here, the following points are worth mentioning:
\begin{itemize}
\item Fermion chains known from QFT have to be observed with the computations for  split() and combine(). 
\item The details of the functions split() and combine() depends on which of these two functions appears first.

\end{itemize}

\subsection{Generation of "Out" Particle/Wave Collection}

The processing of the ia-channel ends with a certain "out" particle/wave combination. With some types of interactions, different "out" particle/wave combinations may occur. The functional model of QT interactions assumes that from the possibly multiple alternative "out" particle/wave combinations only one will actually leave an interaction. This is a further case of "transition from probability to facts" which is possibly not in agreement with standard QFT. 
\footnote{The possible deviation from standard QFT, however, will be difficult to test in experiments.}
The determination of the "out" particle/wave combination may be performed somewhere between step 2 (Formation and processing of ia-channels) and (the  present) Step 3.  (Generation of "Out" Particle/Wave Collection ). In this paper the function is addressed in step 3. The detailed mechanism for the selection of the "out" particle/wave combination is beyond the scope of the present paper. Several alternative mechanisms are imaginable.

After processing the individual ia-channels (in parallel) and dropping those
ia-channels that do not deliver the selected "out" particle/wave combination,
each ia-channel contains the same set of paths, however, with different probability amplitudes for the paths. Therefore, the multiple ia-channels can be (re-) united by the summation of the corresponding amplitudes. According
to QFT rules (usually formulated in terms of Feynman diagrams), the "summation" in some cases has to be performed with a negative sign (i.e., $ amplitude1 - amplitude2 $ instead of $ amplitude1 + amplitude2 $).

\subsubsection{Example: Bhabha Scattering}

Electron-positron scattering may result in lepton pairs (electron, positron), (muon, antimuon), or (tauon, antitauon). For the selection of the "out" particle/wave combination, QFT does not offer any rules besides the equations for the computation of the probability (amplitudes) for the different channels. The selection of the "out" particle/wave combination by the functional model, of course, has to be in accordance with the respective QFT equations.

The summation (or subtraction) of the probability amplitudes also follows the rules defined by QFT.

\section{Related topics}

\subsection{Collapse of the Wave Function} 

When an "out" particle/wave collection is generated, possibly existing "in" particle/wave collections are obsolete. The 
"collapse" of the "in" particle/wave collection consists of two sub-steps:
\begin{itemize}
\item The particles/waves that are  involved in the interaction (i.e., pw1 and pw2) will be discarded from their "in" particle/wave collections.
\item Those  particles/waves that are not involved in the interaction will (of course) survive, however, only the paths which caused the interaction.
\end{itemize}
The fact that this reduction to a single path also affects other particles/waves that are not (directly) involved in the interaction (but "entangled" with the interacting particles/waves) supports entanglement. 

\subsection{Volatile Interaction - Interactions that Do Not Destroy the Superpositions}


Numerous cases are known in the QT concerning
interactions between particles/waves and other matter (particles, atoms, devices, etc.) in which there is no destruction of the superposition before ultimately
a measurement occurs. Typical examples are photons being reflected at mirrors
or photons passing through transparent materials.
Based on the process steps described in section 6., which imply the collapse
of the "in" pw-collections, the question becomes under what circumstances will the mechanism described in section 6.1  be suppressed such that the
"in" particle/wave collections will be preserved. The primary example of this case, which can
hardly be mapped to the process described in section 6, is the interaction
between a particle/wave and a bound system, if the bound system has to be
treated as a single (large) entity. Bound systems are poorly understood with
QFT (see below). This functional model of QT, which is mainly a mapping of
QT/QFT to a process-oriented model, therefore does not yet contain precise criteria for the determination of when the interaction with a bound system will be
volatile. Some widely accepted rules seem to be (1) the higher the energy/mass
of the bound system is, the higher the probability of a volatile interaction becomes;
and (2) the lower the energy/mass of the scattered particle/wave (e.g., photon)
is, the lower the probability of a non-volatile interaction becomes.

The author expects that the development of a better understanding of bound systems
may also provide answers to the question regarding when bound
systems must be treated as an entity (resulting in non-volatile interactions).

\subsection{Interaction with a Bound (State) System}

This functional model of QT interactions is largely derived from the physics
of interactions as provided by QFT. Predictions of the behaviour of bound (state)
systems (henceforth called bound system), such as an atom, a nucleus, or a hadron, can be computed using QFT
for special situations only, and only with considerable effort. QFT includes some
theory and considerations regarding bound systems 
( see  \cite{Weinberg},  \cite{Veltman}, \cite{Griffiths}), 
 however, this does
not include a complete and consistent description of the total system in terms
of QFT constructs such as Feynman diagrams. In 
 \cite{Weinberg} 
S. Weinberg writes, "It must be said that the theory of relativistic effects and radiative corrections
in bound states is not yet in entirely satisfactory shape".

As a consequence of this QFT weakness, the QT functional model cannot yet offer
a complete model of the interaction between a particle/wave and a bound system or
of the interaction between two bound systems. As the major open question, it is not
understood when or to what extent a bound system can/must be treated as an entity
or the whole interaction process has to be broken down to interactions among (elementary)
particles/waves. 

\subsection{Model of Measurement Process}
One of the motivations for the development of the functional model of QT was the author’s suspicion that the apparent nonlinear evolution of the wave function with measurements can get a more plausible explanation, if one relinquishes the requirement that physical processes be describable purely by differential equations. 

The proposed functional model of QT interactions enables a model for the measurement process where there is no explicit "measurement" process (step). Measurement is explained in terms of normal interactions as described in the present paper.
The functional model of measurements described in \cite{diel4} is based on the following assumptions:
\begin{itemize}
\item Measurements require interactions between the measured QT object and part of the measurement apparatus. In general, QT interactions (a) imply transitions from probabilities to facts and (b) have to adhere to the rules and equations of quantum field theory (QFT).
Thus, as described in this paper, the evolution of the wave function during a measurement process is not just a normal linear progression, but a more complicated process which includes the transition from probabilities to facts. 
\item Interactions, in general, support only a limited exchange of information between the "in" QT objects and the "out" QT objects. This limited exchange of information is the cause of some of the peculiarities of QT measurements.
\end{itemize}
With standard QT, the transition of possibilities to facts occurs exclusively
with measurements. The functional model of QT interactions assumes non-deterministic actions at various process steps.

\subsection{Superpositions}
Superposition is one of the key concepts of QT. For the functional model it is assumed that superpositions apply to the paths of the same q-object only, and that superpositions become effective only with interactions (including measurement interactions) occurring at space-time points shared by multiple paths.

\subsection{Entanglement}

As described in section 6, the most important effect of an interaction is the formation of a particle/wave collection with 
multiple paths. 
A path supports correlations and entanglements between the particles/waves leaving the interaction.
Each path represents one of the possible outcomes of an interaction (such as a measurement). As described in section 6.1, the interaction discards all paths of the "in" particle/wave collection except the one selected.

When particles/waves are entangled, the entanglement endures for some time
until it becomes "measured" as a result of an interaction involving one of the entangled particles/waves. "Measured" indicates that due to an interaction one of the entangled particles/waves obtains a definite value for the
measured observable. The measurement of a definite value for particle/wave-1
affects the possible values that can be measured for particle/wave-2. With
the functional model, the determination of a specific measurement value means
selection of a path from a pw-collection. The selection of a path means selection of
a specific value for both particle/wave-1 and particle/wave-2. The path selected
is the only one surviving for the further evolution of particle/wave-2. 

The entanglement concept of the QT functional model maintains the strange
non-locality of QT. However, the non-local effect applies to the elimination of
alternative paths, rather than to direct value changes.

\subsection{Decoherence}

Decoherence theory, like this functional model, discusses process steps such
as the formation of entanglements, interactions, and the (apparent) collapse of
the wave function. Insofar, it may be considered as providing a type of functional
model of QT interactions.

Decoherence is the process of changing the coherent wave function of a local
quantum system through interactions with the environment to a wave function
that is entangled with the environment. H.D. Zeh, one of the founders of 
decoherence theory, calls decoherence an "uncontrolled dis-localization of superpositions" 
(see \cite{ZehPOR}, page 35).

According to Zeh, decoherence occurs constantly as the result of interactions of the observed subsystem with the environment. Even when the observed
subsystem may be isolated, decoherence is unavoidable for the measurement apparatus, and thus, the overall system consisting of the observed subsystem, the
measurement apparatus, and the environment is subject to decoherence 
(see \cite{ZehPOR}, page 34).

Decoherence theory does not assume a collapse of the wave function.  Instead, the many-worlds interpretation 
(see \cite{Everett})
is considered a suitable interpretation of measurements and measurement-like
interactions. The author considers the denial of the collapse of the wave function as an
unavoidable consequence of the insistence on differential equations as the only
physics-conforming way to describe temporal relationships.
In contrast to decoherence theory, the author recognizes a need for an additional
explicit process step besides the unitary evolution defined by the Schr\"odinger
equation. Even with the assumption of a many-worlds interpretation, the "branching" into new worlds (as a replacement of the collapse of the wave function) is
a non-trivial process step that requires further elaboration and (most importantly)
criteria for when it occurs.

\section{Conclusions}

The development of the functional model of QT interactions may be considered
an exercise demonstrating that a functional model of QT that is compatible with
standard QT/QFT is feasible.

Although the functional model aims for maximum compatibility with standard QT/QFT, there are exceptions in the functional model. Examples include some small deviations to standard QFT that have been intentionally included, items where it is not clear what the QT/QFT conformal behavior exactly is, and items where there is no equivalent QT/QFT position because the level of detail is below the scope of QT/QFT. For all these areas, verification by experiments is appropriate.

There is one area in which the functional model is designed to deviate from standard QT/QFT, the concept of the "transition from probabilities to facts". The functional model assumes the transition from probabilities to facts to occur in multiple steps. As indicated within the paper, it will be very difficult to test possible deviations from standard QT in this area.

In addition to the areas in which the functional model intentionally deviates from standard QT/QFT, there may be further accidental deviations. Such accidental deviations should be discovered through computer simulations that compare the predictions of standard QT/QFT with those of the functional model, as described in 
\cite{diel3}.
Preliminary results of these computer simulations show agreement to the extent expected.

The functional model of QT interactions is not considered complete in all
areas. The major areas where the specification is incomplete are (1) the definition
of clear criteria for which interactions do not result in a collapse of the wave
function (see section 7.2) and (2) the treatment of bound systems (see section 7.3).
In both areas, which seem to be interrelated, the existing QT/QFT is not yet in a state that provides much
help for the construction of a functional model.

\end{document}